# Locally Served Network Computers

Jim Gray

February 1995

Technical Report
MSR-TR-95-55

# Locally Served Network Computers
Jim Gray, Microsoft Research, 5 February 1995

Summary: NCs are the natural evolution of PCs, ubiquitous computers everywhere. The current vision of NCs requires two improbable developments: (1) inexpensive high-bandwidth WAN links to the Internet, and (2) inexpensive centralized servers. The large NC bandwidth requirements will force each home or office to have a local server LAN attached to the NCs. These servers will be much less expensive to purchase and manage than a centralized solution. Centralized staff are expensive and unresponsive.

**1. The NC Promise and Premise:** There is enthusiasm for a *Network Computer* (NC) or *Telecomputer*: an inexpensive, stateless device that connects to powerful servers via the Internet. The main benefits of an NC are:
> **Easy to own and use**: it is self installing machines with intuitive interfaces.
> **Inexpensive to own:** it has low capital and maintenance costs.
> **Universal**: it provides access to the entire spectrum of Internet services, as well as personal productivity tools like Word, Excel, and Quicken.

The NC has display, speaker, and printer output devices and keyboard, mouse, and microphone input devices. It has a high-bandwidth connection to the Internet. The NC uses remote applications on the server, or it downloads applications on demand onto a minimal processor and memory within the NC.

The main argument for Network Computers is that **they eliminate the care-and-feeding costs that dominate PC cost-of-ownership**. Studies by Forrester and Gartner estimate that about 75% of a PC's cost-of-ownership is system administration – about 6 k$/year just to support a networked PC user[1]. NC advocates assure us that, since the NC is stateless, NCs will have no care-and-feeding. In essence, all the support cost has moved to the centralized server. Indeed, lower support costs encourage PC customers run diskless workstations today with NetWare or NT servers. X-terminals and diskless workstations have some success in the UNIX world.

**2. A conservative reaction: Yes! But, what's new?** There is little doubt that the telephone, pocket organizer, and gameboy will each evolve into NCs. They will be able to download programs, communicate with others, enhance productivity, or entertain us. The NC is a diskless workstation or X-terminal updated to have a high-quality color display, sound, and a printer. **It is today's multi-media home PC without a disk but with a fast Ethernet connection to an Internet server.**

If you are a technological optimist, the 500$ NC device/price-point will happen, but it will take four to eight years. Right now, such a package has a street price of at least 2000$ (good displays and printers are expensive). The NCs being shown today are actually expensive set-top boxes or low-end Xterms.

**3. Two flaws in the NC argument: (1) Bandwidth, (2) Centralization.** A stand-alone NC is useless. NCs are stateless, so browsers, applets, and data must all be downloaded from the server. A stand-alone NC is not a word processor, not a portable, and not a gameboy. The NC must be connected to the network -- indeed it must be connected to a *fast* network to match PC response times.

The centralization issue is more complex and less technical. NCs postulate a central organization that provides all the management that users now do for themselves. Yes, this will be easier for users, but it will be much more expensive and will be much less responsive. Objective measures show that centralized servers require expensive staff – often this staff is unresponsive to user needs.  I believe customers will be eager to buy self-managed server controlled by the customer – similar to the PBX model.

---
[1] See   Gartner: http://www.gartner.com/hcigdist.htm,
    and    Forrester: http://www.forrester.com/research/cs/1995-ao/jan95csp.html.
Just as a sanity check, it appears Microsoft has about one system administrator or support person per 25 staff and that an Information Technology Group person per 25 staff. That translates to about 10 k$/employee/year in support cost. We average more than one PC per employee ( am told we average seven per employee), but the support cost number is in the right order of magnitude.

Let's consider these the bandwidth and centralization issues in turn:

**3.1. The Bandwidth Problem:** A stateless client-server model (the thin-client model) requires high bandwidth between client and server.   All client applications must be downloaded before they can run, and all the data needs to be downloaded.  Client-server computing has taught us the importance of client caching.   Even across a LAN, it is important for most data and programs to be local to the client after startup.  But, caching creates state that must be managed.  The goal of NCs is to eliminate the management of this state via centralizing the state.  **Automating client state management is an easier, cheaper, and more realistic goal than revolutionizing the WAN business**.

There are two possible WAN bandwidth scenarios over the next few years:
**Telco Armageddon**: A low-cost high-bandwidth telco appears worldwide delivering 10mbps to a hundred million  homes and all offices on the planet by 1998.
**Telco Business as Usual**: Bandwidth out of the house or campus remains a precious commodity in most places.
Business as usual has high probability in the short term (5 years).  We all hope that some new low-cost telco will emerge in the longer term.  In the short term, the WAN-bandwidth problem means that[2]:
- **NCs will not be portable or mobile** because mobile bandwidth is precious and disconnected operation is critical to mobile systems[3].
- **NCs will be served via a LAN to a local server** in the home, SOHO, campus, or site.

If NCs need to be on an Ethernet (or faster) LAN, then each home, school, and office will have a local scaleable super-server and the NCs will hang off its LAN as "terminals."

**3.2. Why Xterms failed and why the issues are different now.**  The arguments for NCs are the same arguments made for Xterms and diskless workstations.   Xterminals and diskless workstations had only modest success.   Why? Why did people buy workstations in preference to Xterms?   Three reasons: (1) Xterminals were not as flexible as PCs – they only did UNIX shells. (2) Xterms were not much cheaper to buy – indeed, the startup cost was higher (an Xterm plus a server rather than a single PC).  (3)  I believe main reason Xterms failed was that PCs give users direct control of their computing resources.   The move to decentralization and a more responsive organization was the real PC driver.

Soon each person will have several PCs or PC equivalents.  Microsoft averages seven PCs per employee.  I manage five and help support several others.   The trend is obvious and frightening. Ubiquitous computing suggests that each of us will soon use and manage dozens of computers. **When there is a PC in every room, the social and organizational barriers against Xterms evaporate** --you do not want a PC per family member per room.  In such a world, the lower management cost becomes the dominant argument for NCs.

**3.3. NCs will fail if they require central servers.**  The popular NC vision postulates stateless NCs attached to giant servers run by a telco or Internet Service Provider – the revenge of the mainframe.  Why do you want to manage your server?  Why not let the phone company do it?  I think there are two answers: convenience and cost.   These two reasons explain why offices have PBXs, why Telco voice-mail has not caught on, and why we avoid doing business with a computer center. If you are old enough,  recall how horrible the data center was, recall how terrible timesharing was, recall how unresponsive the central IS organization was.  Convenience is the main motive, but cost is easier to quantify.

---

[2]  If NCs require both a telecommunications revolution and a computer revolution, they will be gated by the telecom revolution. telcos move very slowly.
[3] Private MicroCell LANs based on radio or infra-red may allow portability within  room or office cells without needing to be tethered to the LAN.  At a minimum, the mobile system will be as stateful as a Newton or Marco PDA system.

I have been looking for data to support these observations. Quantitative evidence is hard to find. In the 1980s, data centers spent 60% on staff and 40% on equipment – very similar to the Forrester study of PC cost-of-ownership today (see footnote 1). In the 1980's there were several rules-of-thumb to estimate staff sizes for a data center. My favorite rules were (1) a systems programmer per MIPS and (2) a data administrator per 10GB. In those days, an IBM MIPS cost a million dollars, ten gigabytes of disk cost a million dollars, and a staff person cost 60k$ per year. So it was good business to hire someone to manage a 1M$ resource. Today the numbers are 8k$ for a CMOS IBM MIPS, 50k$ for ten IBM gigabytes, and 100k$ for a burdened staff person (yes PC MIPS are 1,000x cheaper and PC GBs are 20x cheaper).

Hardware prices have declined by a factor of 300x since 1980. But, people prices for mainframe data centers have not changed a lot. The US Government did a study of its data centers[4]. It examined 205 centers with a total staff of 12,900 people and a 1.2B$ budget. They examined centers of various sizes in detail and found the following statistics:

| Federal Data Center IBM mainframe and compatible Operations Costs | | | | |
|---|---|---|---|---|
| Total MIPS | Average MIPS/machine | Yearly $/MIPS | Full Time Staff / MIPS | Estimated staff/site |
| 1-99 | 36 | $127,500 | .92 | 33 |
| 100-199 | 143 | $129,000 | .51 | 72 |
| 200-299 | 258 | $100,000 | .76 | 196 |
| 300-399 | 336 | $87,600 | .58 | 195 |
| 400 . . . | 767 | $87,700 | .33 | 253 |

These numbers are ASTONISHING. Two hundred people operating a single 340 MIPS system, WOW! Think of the incredible tax these big data centers are charging. *These are just the systems operators and administrators*. These counts do not include applications programmers or computer user.

The government concludes that big data centers get more MIPS per kilo-buck than little data centers. So, on the advice of the data center directors, the government has mandated that each data center should have a minimum of 325 MIPS within two years. This will shrink the number of centers from 205 to about 50 (closing about 150 centers).

I have a 75 MIPS data center in my office (a 25 MIPS laptop and a 50 MIPS tower). Our lab of seven people is a 600 MIPS data center. We sometimes wish for a staff of 200 to manage these machines, but alas we do not have the space or budget for them. So, we daily do the work of 200 systems administrators. Sometimes it feels like that. Actually, NT systems are a LOT easier to install and administer then mainframe MVS systems.

When someone tells you how cheap the mainframe super-servers are going to be, you better hope they are made from GUI/NTclusters that are self-managing. If the NC is talking to the classic mainframe IS/IT structure, you can expect to be paying high prices for the care and feeding of those systems. They will not be less expensive than PCs. If they support thirty users per MIPS, then the data center support costs will be comparable to the per-user PC support costs that NCs are supposed to eliminate.

In addition to being expensive, centralized organizations resist change (e.g., the client upgrade from DOS to Windows 95). Central organizations are generally unresponsive to user needs. This was the main driver for both mini-computers and for PCs.

---

[4] "Consolidation of Federal Data Centers", Prepared by the Federal Data Center Consolidation Committee, John R. Ortago Director, Council of Federal Data Center Directors, Feb. 1995 5203 Leesburgh Pike, Suite 400, Falls Church, VA 22041, (703-756-4111)

**4. A better NC model: a plug-and-play to a local server:** NCs are an evolution of the PC world. Today the PC, the printer, the portable, and the tower are each an independent computer. Each requires some setup and some management. It would be much simpler if the OS software automated this making all these computers look like a single management entity. I should be able to just plug computers and applications into the network, much as I plug telephones in today.

The bandwidth requirements suggest that each home, classroom and each personal office will need a local server – all the NCs and ubiquitous computers in the area will be LAN connected to this server. You will be able to hang as many devices off this server as you like. In your office, the server will have an infra-red LAN to talk to all the devices there: your portables, the printers, the phones,… In the home, the server will drive all the NCs spread throughout the house in place of telephones, radios, TVs, intercoms, and gameboys. The server will store your data and programs, give you a single thing to manage, and it will optimize your use of the expensive telco resource at the curb. This is an evolutionary path where the current PCs evolve into servers managing all the intelligent devices about to come into our lives.

This home and group server will require management. The server software will manage everything that can be automated. But it will need guidance on policy issues like security, quotas, and billing. In this world, the distinction between NCs and PCs is that NCs cannot operate detached from the network, while PCs may be used in stand-alone (e.g. mobile) devices. Otherwise, NCs and PCs will be equally easy to manage since the software will automate the management tasks we perform today.

**5. Summary**: Stateless simple computers, Network Computers, are coming. But they will be LAN connected to local servers. By caching data and programs at the server, this architecture minimizes system management and optimizes use of expensive WAN bandwidth. The distinction between NCs and PCs is that PCs support stand-alone (e.g., mobile) operation.

**Acknowledgments:** This paper developed from discussions with Gordon Bell, Butler Lampson, and Phil Garrett. Rick Rashid's skepticism about support costs challenged me to find quantitative arguments.